\documentclass[12pt]{article}
\usepackage{amsmath,amssymb,amsthm,cite}
\usepackage{fullpage}

\InputIfFileExists{whatnot.tex}{}{}

\newcommand{\eps}{\epsilon}        
\newcommand{\F}{\mathcal{F}}       
\newcommand{\Ga}{\Gamma}           
\renewcommand{\H}{\mathcal{H}}     
\newcommand{\half}{\tfrac{1}{2}}   
\newcommand{\Hom}{\mathrm{Hom}}    
\newcommand{\id}{\mathrm{id}}      
\newcommand{\K}{\mathcal{K}}       
\newcommand{\Km}{\mathcal{K}_-}    
\newcommand{\Kp}{\mathcal{K}_+}    
\newcommand{\Krm}{\mathrm{K}}      
\newcommand{\ket}[1]{|#1\rangle}   
\newcommand{\ox}{\otimes}          
\newcommand{\Rbar}{{\overline{R}}} 
\newcommand{\sepword}[1]{\quad\mbox{#1}\quad} 
\newcommand{\upper}{{\mathrm{up}}} 
\newcommand{\vf}{\varphi}          
\newcommand{\vft}{\tilde{\varphi}} 
\def\<#1,#2>{\langle#1,#2\rangle}  

\makeatletter
\def\section{\@startsection{section}{1}{\z@}{-3.5ex plus -1ex minus
      -.2ex}{2.3ex plus .2ex}{\large\bf}}
\def\subsection{\@startsection{subsection}{2}{\z@}{-3.25ex plus -1ex
      minus -.2ex}{1.5ex plus .2ex}{\normalsize\bf}}
\makeatother

\addtocounter{MaxMatrixCols}{2} 

\newcommand{\fish}{\parbox{1.5pc}{\begin{picture}(10,10) 
\put(1,5){\qbezier(0,-2)(8,10)(16,-2)}
\put(1,5){\qbezier(0,2)(8,-10)(16,2)}
\end{picture}}}

\newcommand{\bikini}{\parbox{2pc}{\begin{picture}(10,5) 
\put(0,2){\qbezier(0,-1.5)(6,7.5)(11.4,0)}
\put(0,2){\qbezier(0,1.5)(6,-7.5)(11.4,0)}
\put(10.8,2){\qbezier(0.6,0)(6,7.5)(12,-1.5)}
\put(10.8,2){\qbezier(0.6,0)(6,-7.5)(12,1.5)}
\end{picture}}}

\newcommand{\winecup}{\parbox{1.4pc}{\begin{picture}(10,10) 
\put(5.9,-1.4){\line(2,3){9}}
\put(11.1,-1.4){\line(-2,3){9}}
\put(4,7.5){\qbezier(0,1.8)(4.5,5)(9,1.8)}
\put(4,7.5){\qbezier(0,1.8)(4.5,-1.25)(9,1.8)}
\end{picture}}}

\newcommand{\trikini}{\parbox{3pc}{\begin{picture}(5,5) 
\put(0,2){\qbezier(0,-1.5)(6,7.5)(11.4,0)}
\put(0,2){\qbezier(0,1.5)(6,-7.5)(11.4,0)}
\put(10.8,2){\qbezier(0.6,0)(6,7.5)(12,0)}
\put(10.8,2){\qbezier(0.6,0)(6,-7.5)(12,0)}
\put(22,2){\qbezier(0.6,0)(6,7.5)(12,-1.5)}
\put(22,2){\qbezier(0.6,0)(6,-7.5)(12,1.5)}
\end{picture}}}

\newcommand{\shark}{\parbox{2pc}{\begin{picture}(10,10) 
\put(0,6){\qbezier(0,-1.5)(6,7.5)(11.4,0)}
\put(0,6){\qbezier(0,1.5)(6,-7.5)(11.4,0)}
\put(11.4,6){\line(2,1){10}}
\put(11.4,6){\line(2,-1){10}}
\put(19.4,2){\qbezier(0,0)(-4,4)(0,8)}
\put(19.4,2){\qbezier(0,0)(4,4)(0,8)}
\end{picture}}}

\newcommand{\stye}{\parbox{2pc}{\begin{picture}(10,15) 
\put(1.4,7){\qbezier(0,-2.8)(11.2,14)(22.4,-2.8)}
\put(1.4,7){\qbezier(0,2.8)(11.2,-14)(22.4,2.8)}
\put(12.6,12){\circle{8}}
\end{picture}}}

\newcommand{\tetra}{\parbox{1.75pc}{\begin{picture}(10,20) 
\put(-2,10){\line(1,0){10}}
\put(12,10){\line(1,0){10}}
\put(0,10){\line(1,1){10}}
\put(0,10){\line(1,-1){10}}
\put(10,-2){\line(0,1){24}}
\put(20,10){\line(-1,1){10}}
\put(20,10){\line(-1,-1){10}}
\end{picture}}}

\newcommand{\twoscoops}{\parbox{2.3pc}{\begin{picture}(10,10) 
\put(2,10){\qbezier(0,0)(6,6)(12,0)}
\put(2,10){\qbezier(0,0)(6,-6)(12,0)}
\put(14,10){\qbezier(0,0)(6,6)(12,0)}
\put(14,10){\qbezier(0,0)(6,-6)(12,0)}
\put(14,0){\qbezier(-14,12)(-8,3)(2,-2)}
\put(14,0){\qbezier(-2,-2)(8,3)(14,12)}
\end{picture}}}

\newcommand{\roll}{\parbox{1.5pc}{\begin{picture}(10,15) 
\put(9,8){\circle{16}}
\put(9,8){\qbezier(-8,-8)(0,0)(-8,8)}
\put(9,8){\qbezier(8,-8)(0,0)(8,8)}
\end{picture}}}

\newcommand{\kite}{\parbox{1.7pc}{\begin{picture}(20,15) 
\put(7,0){\line(-1,3){6.4}}
\put(7,0){\line(1,3){4.8}}
\put(5,-1.5){\line(4,3){14}}
\put(11.8,14.4){\line(2,-3){7.2}}
\put(2.2,14.4){\qbezier(0,0)(4.8,4.8)(9.6,0)}
\put(2.2,14.4){\qbezier(0,0)(4.8,-4.8)(9.6,0)}
\end{picture}}}

\newcommand{\catseye}{\parbox{1.9pc}{\begin{picture}(10,15) 
\put(0,7){\qbezier(0,-2.8)(11.2,14)(22.4,-2.8)}
\put(0,7){\qbezier(0,2.8)(11.2,-14)(22.4,2.8)}
\put(11.2,1.4){\qbezier(0,0)(-4,5.6)(0,11.2)}
\put(11.2,1.4){\qbezier(0,0)(4,5.6)(0,11.2)}
\end{picture}}}

\newcommand{\sunset}{\parbox{1.5pc}{\begin{picture}(0,12) 
\put(9,6){\circle{12}}
\put(0,6){\line(1,0){18}}
\end{picture}}}

\begin{document}

\title{\bf Combinatorics of renormalization\\
as matrix calculus}

\author{
Kurusch Ebrahimi-Fard
\\
Physics Institute, Bonn University,
\\
Nussallee 12, Bonn 53115, Germany
\\[0.2cm]
Jos\'e M. Gracia-Bond\'{\i}a
\\
Departamento de F\'{\i}sica Te\'orica I, Universidad Complutense,
\\
Madrid 28040, Spain
\\[0.2cm]
Li Guo
\\
Department of Mathematics and Computer Science, Rutgers University,
\\
Newark, NJ 07102, USA
\\[0.08cm]
and
\\[0.08cm]
Joseph C. V\'arilly
\\
Departamento de Matem\'aticas, Universidad de Costa Rica,
\\
San Jos\'e 2060, Costa Rica}

\date{}

\maketitle

\vspace{1.5pc}

\begin{abstract}
We give a simple presentation of the combinatorics of renormalization
in perturbative quantum field theory in terms of triangular matrices.
The prescription, that may be of calculational value, is derived from
first principles, to wit, the ``Birkhoff decomposition'' in the
Hopf-algebraic description of renormalization by Connes and Kreimer.
\end{abstract}

\vspace{1.5pc}

\noindent
2001 PACS Classification: 03.70.+k, 11.10.Gh, 02.10.Hh, 02.10.Ox


\newpage

\section{Introduction}

A Hopf algebra structure underlying the combinatorics of perturbative
renormalization was recognized by Kreimer~\cite{KreimerOriginal}. Some
have worried about the practical usefulness of his insight for
organizing everyday computations in quantum field theories. A partial
answer to this legitimate question is given in~\cite{BK}, for
instance. This paper presents another partial answer, in a different
vein. We show that the Feynman rules collectively possess a triangular
matrix representation, such that the renormalization map becomes a
matrix operation. This infinite matrix can be truncated almost
\textit{ad libitum}.

The procedure is general, largely independent of the renormalization
scheme (although we illustrate everything with the MS-scheme in
dimensional regularization), and essentially independent of the
particular field theory model one works with. The latter enters just
in filling up the matrix entries, generally entailing further
simplification. Only knowledge of linear algebra and quantum fields
are required; no Hopf algebra is (openly) used. More detail is found
in~\cite{Ausonia}.

First we recall the algebraic behaviour of the subtraction
map~$K$: the whole paper turns around its Rota--Baxter property.
In Section~3 $K$ is lifted to the matrix level, and the
computational recipe for the matrix counterterm and matrix
renormalization maps is found. In the following section we verify
that this reproduces the diagrammatic Bogoliubov operation; we
take examples from the~$\phi^4_4$~model, and compare with the
tables in~\cite{KleinertSF}. In Section~5 we rework the matrix
representation using the map~$K_+$ that picks out the finite
parts. The next two sections contain mathematical summaries. In
Section~7 we sketch our derivation of the matrix representation.
Finally we examine the outlook.

\section{The subtraction map as a Rota--Baxter operator}

Consider Laurent series
\begin{equation}
S(\eps) = \frac{a_{-n}}{\eps^n} + \frac{a_{-n+1}}{\eps^{n-1}}
+\cdots+ \frac{a_{-1}}{\eps} + a_0 + a_1\eps + \cdots \,.
\label{eq:ya-empezamos}
\end{equation}
With the ordinary multiplication, they form a commutative algebra~$V$
with unit. Consider further the operation $K$ which picks out the pure
pole part
$$
K[S](\eps) = \frac{a_{-n}}{\eps^n} + \frac{a_{-n+1}}{\eps^{n-1}}
+\cdots+ \frac{a_{-1}}{\eps},
$$
and the operation $K_+ := \id - K$ keeping the finite part,
$$
K_+[S](\eps) = a_0 + a_1\eps + \cdots \,.
$$
The projector condition $K^2 = K$ ensures that the intersection
between $K(V)$ and $K_+(V)$ is zero. The product of two elements
of~$K(V)$ remains in $K(V)$ ---and likewise for $K_+(V)$. The key
property
\begin{equation}
K[S_1]\,K[S_2] = K\bigl[ K[S_1]S_2 + S_1K[S_2] - S_1S_2 \bigr],
\label{eq:Rota-Baxter}
\end{equation}
is easy to check ---see Section~6. It makes $K$ a
\textit{Rota--Baxter} operator~\cite{EbrahimiFGuoKr}; also $K_+$ is a
Rota--Baxter operator. All this applies in particular to series
corresponding to dimensionally regularized integrals in the MS-scheme
subtraction. Our arguments are purely combinatorial, so we need not
worry about the precise form of the $a_i$ coefficients. We adopt for
the subtraction operator $K$ the notation of \cite{CaswellK}, followed
in \cite{KleinertSF}; sometimes we write $K_-$ for clarity.

\section{Setting up the recipe}
\label{sec:setup}

In this section we suppose given the pair $(V,K)$ of a commutative
algebra with unit and a Rota--Baxter projector, first as an abstract
framework; we always have in mind the algebra of unrenormalized
Feynman amplitudes of the form~\eqref{eq:ya-empezamos}. Consider upper
triangular matrices of finite size with entries in~$V$, of two types:
nilpotent, i.e., with 0's on the diagonal, and unipotent, i.e., with
1's on the diagonal:
$$
Z = \begin{pmatrix}
0       & *      & \dots  & \dots     & * \\
0       & 0      & \ddots & Z_{ij}    & \vdots \\
\vdots  & \ddots & \ddots & \ddots    & \vdots \\
\vdots  & \ddots & \ddots & \ddots    & * \\
0       & \ldots & \ldots & 0         & 0 \end{pmatrix};
\qquad
\vf = \begin{pmatrix}
1       & *      & \dots  & \dots     & * \\
0       & 1      & \ddots & \vf_{ij}  & \vdots \\
\vdots  & \ddots & \ddots & \ddots    & \vdots \\
\vdots  & \ddots & \ddots & \ddots    & * \\
0       & \ldots & \ldots & 0         & 1 \end{pmatrix}.
$$
We define Rota--Baxter operations $\Km = \K$ and $\Kp$, on algebras
$M^\upper(V)$ of upper triangular matrices with scalar diagonals and
with entries in~$V$, by extending the maps~$K$ and~$K_+$
componentwise,
$$
(\K[\vf])_{ij}  := K[\vf_{ij}],    \quad
(\Kp[\vf])_{ij} := K_+[\vf_{ij}].
$$
Verification for $\K$, $\Kp$ of the analogue of~\eqref{eq:Rota-Baxter}
is immediate; but the algebras $M^\upper(V)$ are no longer
commutative. We seek to factorize an arbitrary unipotent element $\vf$
in the form
\begin{equation}
\vf = \vf_+ \, \vf_-^{-1},
\label{eq:mano-de-la-princesa}
\end{equation}
where the factors $\vf_- \in 1 + \K[M^\upper(V)]$, $\vf_+ \in
\Kp[M^\upper(V)]$ are also unipotent; note that they are unique. This
can be called a matrix Birkhoff decomposition.

If $\K[\log\vf]$ and $\Kp[\log\vf]$ happened to commute, it would be
enough to choose $\vf_+ = e^{\Kp[\log\vf]}$ and
$\vf_- = e^{-\K[\log\vf]}$. In general, that is not so; but we are
able to compensate for the lack of commutativity between the images of
$\Km$ and~$\Kp$. For that, consider the equations
$$
\vf_-      = 1 - \Km\bigl[ (\vf - 1)\vf_- \bigr]  \sepword{and}
\vf_+^{-1} = 1 - \Kp\bigl[ \vf_+^{-1}(\vf - 1) \bigr],
$$
respectively solved by
\begin{align}
\vf_-      &= 1 - \Km[\vf - 1] + \Km[(\vf - 1)\Km[\vf - 1]] -\cdots;
\label{eq:phi-factor-minus}
\\
\vf_+^{-1} &= 1 - \Kp[\vf - 1] + \Kp[\Kp[\vf - 1](\vf - 1)] -\cdots.
\label{eq:phi-factor-plusinv}
\end{align}
Both series terminate. Atkinson's theorem~\cite{Atkinson} asserts that
these matrices $\vf_-$, $\vf_+^{-1}$
verify~\eqref{eq:mano-de-la-princesa}. The proof runs as follows:
\begin{align*}
\vf_+^{-1} \vf_-
&= 1 - \Kp\bigl[ \vf_+^{-1}(\vf - 1) \bigr]
- \Km\bigl[ (\vf - 1)\vf_- \bigr]
+ \Kp\bigl[ \vf_+^{-1}(\vf - 1)\bigr] \Km\bigl[(\vf - 1)\vf_- \bigr]
\\
&= 1 - \vf_+^{-1}(\vf - 1)\vf_-,
\end{align*}
after some work with the Rota--Baxter property.

The matrix $\vf_+$ is what we are really after. It can obviously be
obtained as $\vf\vf_-$, or by inverting~\eqref{eq:phi-factor-plusinv},
using the geometric series formula
$$
\vf_+ = 1 - (\vf_+^{-1} - 1) +  (\vf_+^{-1} - 1)^2 - \cdots.
$$
A better course is perhaps to observe that, by the same token as in
\eqref{eq:phi-factor-minus}, we obtain
$$
\vf_+ = 1 - \Kp\bigl[ (\vf^{-1} - 1)\vf_+ \bigr].
$$
Thus, the respective formulas for the components of $\vf_-$
and~$\vf_+$ are
\begin{align}
{\vf_-}_{ij} &= - K_-[\vf_{ij}] +
\sum_{k=1}^{j-i-1} \sum_{i<l_1<l_2<\cdots<l_k<j} (-)^{k-1}
K_-\bigl[\vf_{il_1}\, K_-[\vf_{l_1l_2} \dots K_-[\vf_{l_kj}]\dots]\bigr],
\label{eq:war-horse}
\\
{\vf_+}_{ij} &= - K_+[\vf_{ij}^{-1}]
+ \sum_{k=1}^{j-i-1} \sum_{i<l_1<l_2<\cdots<l_k<j} (-)^{k-1}
K_+\bigl[\vf_{il_1}^{-1}\, K_+[\vf_{l_1l_2}^{-1} \dots
K_+[\vf_{l_kj}^{-1}]\dots]\bigr].
\nonumber
\end{align}
These similar formulas are our workhorses; with the appropriate
definition of~$\vf$, the matrix~$\vf_-$ will be seen to contain all
the information on counterterms in renormalization; and~$\vf_+$ on the
renormalized quantities.

\section{Making the recipe work}
\label{sec:recipe}

Now we make explicit how the Feynman rules specify such
operators~$\vf$. Recall that if $\Ga_i \subseteq \Ga_j$ is a
superficially divergent subgraph of~$\Ga_j$, the
\textit{cograph}~$\Ga_j/\Ga_i$ is obtained by shrinking~$\Ga_i$ to a
vertex within~$\Ga_j$. We only consider subgraphs that are generalized
vertices~\cite{CaswellK}. Chosen an $n$-point function, the spaces of
vectors on which the matrices act are spanned by the corresponding
(superficially divergent, \textit{connected}, amputated) Feynman
graphs. We may use the familiar bra-ket notation to denote the
diagrams as vectors. A basis $\ket{\Ga_1}, \ket{\Ga_2},
\ket{\Ga_3},\dots$ for such a space can be ordered in many ways, the
only conditions being that $\ket{\Ga_1} = \ket{\emptyset}$ ---the
empty diagram--- and that each cograph of any~$\Ga_l$ occurs in the
basis as some $\Ga_m$ with $m < l$. It is then convenient to order the
basis by number of loops (or vertices, if we work on coordinate
space); but the order within a given loop-number sector is immaterial.
Once the external structure and the basis are fixed, we fill up the
entries of a matrix by the rule: for $i \neq j$,
$$
\vf_{ij} = \sum_{\Ga'} \mbox{(unrenormalized) amplitude of $\Ga'$}
\sepword{if}  \Ga_i \simeq \Ga_j/\Ga' \,,
$$
otherwise $\vf_{ij} = 0$. This entails triangularity, since
$\vf_{ij} = 0$ if $i > j$. We set $\vf_{ii} = 1$ for all~$i$. Note
that $\Ga'$ need not belong to the basis list (it might be
disconnected, for one thing). Let $\vft(\Ga')$ be the unrenormalized
amplitude of~$\Ga'$. We just said that the coefficient of
$\ket{\Ga_i}$ in $\vf(\ket{\Ga_j})$ is $\sum_{\Ga'}\vft(\Ga')$ for
$\Ga_i \simeq \Ga_j/\Ga'$. The notation is appropriate because $\vft$
is the abstract object represented by the matrix $\vf$ (see
Section~7); it has the property that
$$
\vft(\Ga_i\Ga_j) := \vft(\Ga_i \cup \Ga_j) = \vft(\Ga_i)\,\vft(\Ga_j).
$$
This property is shared by the elements of $\vf_-$ and~$\vf_+$. As a
bonus, it allows to simplify the notation later on, by simply
omitting~$\vft$.

When we truncate the matrices we are not obliged to include all the
diagrams belonging to the higher sector ---and we can also choose, for
whatever purpose, particular classes of diagrams, subject to the
aforementioned two conditions. With that, the first row of~$\vf$ is
given by $1,\vft(\Ga_2),\vft(\Ga_3),\dots$, the unrenormalized
amplitudes of all the diagrams; and the analogously defined first row
$(1,\vft_-(\Ga_2),\vft_-(\Ga_3),\dots)$ of~$\vf_-$ will yield all the
counterterms of the theory!

We take as a simple \textit{example} the space of graphs relevant to
the 4-point function for the $\phi^4_4$ model, truncated to the 12
tadpole-free diagrams (including the empty one) up to three loops. We
adopt the order of~\cite{KleinertSF} for the basis, as follows:
$$
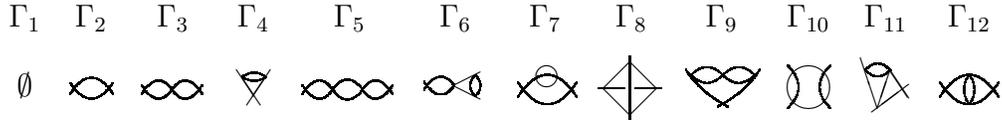
\begin{matrix}
\Ga_1 & \Ga_2 & \Ga_3 & \Ga_4
& \Ga_5 & \Ga_6 & \Ga_7 & \Ga_8
& \Ga_9 & \Ga_{10} & \Ga_{11} & \Ga_{12}
\\[3\jot]
\emptyset & \fish & \bikini & \winecup
& \trikini & \shark & \stye & \tetra
& \twoscoops & \roll & \kite & \catseye
\end{matrix}
$$
Notice that $\Ga_2 = \Ga_7/\Ga'$ where the sunset diagram
$\Ga' = \sunset$ does not appear in the basis list. We find that $\vf$
is equal to
\bgroup
\footnotesize \setlength{\arraycolsep}{3.0pt} 
$$
\begin{pmatrix}
1 & \vft(\Ga_2) & \vft(\Ga_3) & \vft(\Ga_4) & \vft(\Ga_5)
& \vft(\Ga_6) & \vft(\Ga_7) & \vft(\Ga_8) & \vft(\Ga_9)
& \vft(\Ga_{10}) & \vft(\Ga_{11}) & \vft(\Ga_{12})
\\
& 1 & 2\vft(\Ga_2) & \vft(\Ga_2) & 2\vft(\Ga_3) + \vft^2(\Ga_2)
& \vft(\Ga_4) + \vft^2(\Ga_2) & \vft(\Ga') & 0 & \vft(\Ga_3)
& \vft^2(\Ga_2) & \vft(\Ga_4) & 2\vft(\Ga_4)
\\
&& 1 & 0 & 3\vft(\Ga_2) & \vft(\Ga_2) & 0 & 0 & 0 & 0 & 0 & \vft(\Ga_2)
\\
&&& 1 & 0 & \vft(\Ga_2) & 0 & 0 & 2\vft(\Ga_2)
& 2\vft(\Ga_2) & \vft(\Ga_2) & 0
\\
&&&& 1 & 0 & 0 & 0 & 0 & 0 & 0 & 0 \\
&&&&& 1 & 0 & 0 & 0 & 0 & 0 & 0 \\
&&&&&& 1 & 0 & 0 & 0 & 0 & 0 \\
&&&&&&& 1 & 0 & 0 & 0 & 0 \\
&&&&&&&& 1 & 0 & 0 & 0 \\
&&&&&&&&& 1 & 0 & 0 \\
&&&&&&&&&& 1 & 0 \\
&&&&&&&&&&& 1 \end{pmatrix}
$$
\egroup The abundance of zeros is welcome in the calculation: with
this concrete form of~$\vf$ the series~\eqref{eq:phi-factor-minus}
stops after three iterations. Omitting $\vft$ from the notation as
advertised, one reads off from~\eqref{eq:war-horse} the first row of
$\vf_-$, containing the counterterms for the eleven nontrivial
diagrams, that we write as the column matrix $\vf^T_{-j1}$ in the
following display. These expressions for $\vft_-(\Ga_j)$ coincide with
those listed in the tables in~\cite{KleinertSF}, where they are
denoted $-K\Rbar(\Ga_j)$. But note that here the Bogoliubov
preparation map $\Rbar$ does not appear explicitly; our result, with
one exception, is not recursively presented, and, with the help of
some symbolic programming, can be obtained at one stroke. It is clear
that the method will jointly handle large numbers of multiloop
diagrams with ease. (Of course, we are not claiming that it is always
quicker than the standard procedures.)
$$
\begin{pmatrix}
1 \\[2\jot] \vft_-\bigl(\fish\bigr) \\[2\jot]
\vft_-\bigl(\bikini\bigr) \\[2\jot]
\vft_-\bigl(\winecup\bigr) \\[2\jot]
\vft_-\bigl(\trikini\bigr) \\[2\jot]
\vft_-\bigl(\shark\bigr) \\[2\jot]
\vft_-\bigl(\stye\bigr) \\[2\jot]
\vft_-\bigl(\tetra\bigr) \\[2\jot]
\vft_-\bigl(\twoscoops\bigr) \\[2\jot]
\vft_-\bigl(\roll\bigr) \\[2\jot]
\vft_-\bigl(\kite\bigr) \\[2\jot]
\vft_-\bigl(\catseye\bigr)
\end{pmatrix}
= \begin{pmatrix}
1 \\[2\jot]
- K[\fish] \\[2\jot]
  K[\fish]\,K[\fish] \\[2\jot]
- K[\winecup] + K[\fish K[\fish]] \\[2\jot]
- K[\fish]\,K[\fish]\,K[\fish] \\[2\jot]
- K[\fish]\,\vft_-\bigl(\winecup\bigr) \\[2\jot]
- K[\stye] + K[\fish K[\sunset]] \\[2\jot]
- K[\tetra] \\[2\jot]
- K\bigl[\twoscoops\bigr] + 2 K[\winecup K[\fish]]
- K\bigl[\fish\bigl(K[\fish]\bigr)^2\bigr]
\\[2\jot]
- K\bigl[\roll\bigr] + 2 K[\winecup K[\fish]]
- K\bigl[\fish\bigl(K[\fish]\bigr)^2\bigr]
\\[2\jot]
- K\bigl[\kite\bigr] + K[\fish]\,K[\winecup] + K[\fish\winecup]
- [K[\fish]]^{\{3\}}
\\[2\jot]
- K[\catseye] + K[\fish^2 K[\fish]]  + 2 K[\fish K[\winecup]]
- 2 [K[\fish]]^{\{3\}}
\end{pmatrix}.
$$
Above we wrote $[K[\fish]]^{\{3\}} := K[\fish\,K[\fish\,K[\fish]]]$.
The graphs $\Ga_3,\Ga_5,\Ga_6$ are cutvertex, for which~$\vft$ and the
renormalization map are known to factorize. For
them~\eqref{eq:war-horse} \textit{prima facie} gives a more
complicated expression, that can be reduced to the expressions shown
by some Rota--Baxter gymnastics. We give the example of
$\vft_-\bigl(\shark\bigr)$. From our matrix operations:
\begin{align*}
\vft_-\bigl(\shark\bigr)
&= - K\bigl[\shark\bigr] + K\bigl[ \winecup K[\fish] \bigr]
+ K\bigl[ \fish K\bigl[\winecup\bigr] \bigr]
\\
&\qquad + K\bigl[\bikini K[\fish] \bigr]
+ K\bigl[\fish K[\fish^2] \bigr] - 3 [K[\fish]]^{\{3\}}.
\end{align*}
Since $K\bigl[\shark\bigr] = K\bigl[\fish \winecup\bigr]
= - K[\fish]\,K\bigl[\winecup\bigr] + K\bigl[K[\fish] \winecup\bigr]
+ K\bigl[K\bigl[\winecup\bigr] \fish\bigr]$, we get
\begin{align*}
\vft_-\bigl(\shark\bigr)
&= K[\fish]\,K\bigl[\winecup\bigr] + K\bigl[\fish^2 K[\fish]\bigr]
+ K\bigl[\fish K[\fish^2]\bigr] - 3 [K[\fish]]^{\{3\}}
\\
&= K[\fish]\,K\bigl[\winecup\bigr] + K[\fish^3]
+ K[\fish^2]\,K[\fish] - 3 [K[\fish]]^{\{3\}}.
\end{align*}
To continue, we invoke the classical Bohnenblust--Spitzer
identity~\cite{EbrahimiFGuoEofM}:
$$
n!\,[K[A]]^{\{n\}} := n!\,\underbrace{K\bigl[A K[A\dots
K[A]\dots]\bigr]}_{n\;\mathrm{times}} = \sum_{P\in\, \Pi_n}
\prod_{p\in\, P} (|p| - 1)! K(A^{|p|}),
$$
which is itself derivable from the Rota--Baxter identity; here $\Pi_n$
is the set of partitions $p$ of the set $\{1,\dots,n\}$. For the
present case,
$$
6 [K[\fish]]^{\{3\}} = K[\fish]\,K[\fish]\,K[\fish]
+ 3 K[\fish^2]\,K[\fish] + 2 K\bigl[\fish^3\bigr],
$$
implying that
\begin{align*}
&\vft_-\bigl(\shark\bigr) = K[\fish]\,K\bigl[\winecup\bigr] - \half
K[\fish]\,K[\fish]\,K[\fish] - \half K[\fish]\,K[\fish^2] =
K[\fish]\,K\bigl[\winecup\bigr]
\\
&- K[\fish]\,K[\fish K[\fish]] = \bigl(-K[\fish]\bigr)\,
\bigl(-K\bigl[\winecup\bigr] + K[\fish K[\fish]]\bigr) =
\vft_-\bigl(\fish\bigr) \, \vft_-\bigl(\winecup\bigr).
\end{align*}
According to the theory underlying the matrix representation
(Section~7), the factorization property of~$\vft_-$ is automatic: we
did not \textit{prove} anything, just performed an internal check.

\section{The matrix representation in terms of $K_+$}

Renormalization theory is usually formulated in terms of
subtractions~\cite{KleinertSF,CaswellK,Collins}. For good reasons:
for instance, in the MS-scheme the counterterms are local,
independent of the mass and the renormalization scale; this helps
to establish renormalization group equations. Also, the derivation
of the whole procedure is simpler in terms of subtractions.
However, the understanding of renormalization as an approximation
process, rather than a cancellation of infinities, is thereby
lost. Recently 't~Hooft expressed the \textit{desideratum} of a
renormalization scheme exclusively containing dressed
vertices~\cite{tHooftFinite}. We take here a step in this
direction by rephrasing the renormalization of (regularized)
Feynman graphs with subdivergences in terms of~$K_+$. The magic of
the triangular matrix representation implies that the matrix
$\vf_+$ must give, graph by graph, the completely renormalized
expressions $(\id - K)\Rbar(\Ga) = K_+\Rbar(\Ga)$. Inspection of
the equations in Section~\ref{sec:setup}, on the other hand, shows
that the same calculation method for $\vf_-$ in terms of $K_-$
yields $(1,\vft_+(\Ga_2),\vft_+(\Ga_3), \dots)$ in terms of~$K_+$,
provided one starts by inverting the matrix~$\vf$. This is a small
price to pay, and again a bit of symbolic programming goes a long
way. We illustrate the procedure with the same model example:
\begin{equation*}
\begin{pmatrix}
\vft_+\bigl(\bikini\bigr) \\[1\jot]
\\[-0.4cm]
\vft_+\bigl(\winecup\bigr) \\[1\jot]
\\[-.4cm]
\vft_+\bigl(\trikini\bigr) \\[1\jot]
\\[-.4cm]
\vft_+\bigl(\shark\bigr) \\[1\jot]
\\[-.3cm]
\vft_+\bigl(\stye\bigr) \\[1\jot]
\\[.1cm]
\vft_+\bigl(\twoscoops\bigr) \\[1\jot]
\\[2\jot]
\vft_+\bigl(\roll\bigr) \\[1\jot]
\\[2\jot]
\vft_+\bigl(\kite\bigr) \\[1\jot]
\\[2\jot]
\vft_+\bigl(\catseye\bigr)\\
\\
\end{pmatrix}
=
\begin{pmatrix}
K_+[\fish]\,K_+[\fish]
\\[2\jot]
K_+\bigl[\winecup\bigr] + K_+[\fish]\,K_+[\fish] - K_+[\fish
K_+[\fish]]
\\[2\jot]
K_+[\fish]\,K_+[\fish]\,K_+[\fish]
\\[2\jot]
K_+[\fish]\,\bigl(K_+\bigl[\winecup\bigr] +
K_+[\fish]\,K_+[\fish]-K_+[\fish K_+[\fish]]\bigr)
\\[2\jot]
K_+\bigl[\stye\bigr] + K_+[\fish]\,K_+[\sunset] - K_+\bigl[\sunset
K_+[\fish]\bigr]
\\[2\jot]
\left\{\begin{aligned} & K_+\bigl[\twoscoops\bigr] + 2
K_+\bigl[\winecup\bigr]\,K_+[\fish] - K_+[\fish]\,K_+[\fish^2]
\\[\jot]
& \quad - 2 K_+\bigl[\fish K_+\bigl[\winecup\bigr]\bigr] -
K_+[\fish^2 K_+[\fish]] + 2 [K_+[\fish]]^{\{3\}}
\end{aligned}\right\}
\\[6\jot]
\left\{\begin{aligned}
& K_+\bigl[\roll\bigr] + 2 K_+\bigl[\winecup\bigr]\,K_+[\fish]
- K_+[\fish]\,K_+[\fish^2]
\\[\jot]
& \quad - 2 K_+\bigl[\fish K_+\bigl[\winecup\bigr]\bigr] - K_+[\fish^2
K_+[\fish]] + 2 [K_+[\fish]]^{\{3\}}
\end{aligned}\right\}
\\[6\jot]
\left\{\begin{aligned}
& K_+\bigl[\kite\bigr] + K_+\bigl[\winecup\bigr]\,K_+[\fish]
- K_+[\fish]\,K_+[\fish^2]
\\[\jot]
& \quad - K_+\bigl[\winecup\fish\bigr] + [K_+[\fish]]^{\{3\}}
\end{aligned}\right\}
\\[6\jot]
\left\{\begin{aligned}
& K_+\bigl[\catseye\bigr] + 2 K_+\bigl[\winecup\bigr]\,K_+[\fish]
- K_+[\fish]\,K_+[\fish^2]
\\[\jot]
& \quad - 2K_+\bigl[\winecup K_+[\fish]\bigr] +
K_+\bigl[\fish\bigl(K_+[\fish]\bigr)^2\bigr]
\end{aligned}\right\}
\end{pmatrix}.
\end{equation*}
The simpler cases have been omitted. The reader is reminded
that~$K_+[\fish]$ actually means $K_+[\vft(\fish)]$, and so on.

\section{More on Rota--Baxter operators}

A few extra comments on the Rota--Baxter property of $K_-$ and $K_+$
are in order. In the work by Kreimer, the Rota--Baxter property
appears for the first time in~\cite{KreimerChenIter}, under the name
``multiplicativity constraint''. The maps $K_-$, $K_+$ may be regarded
as generalized integrals. Indeed, let us insert a parameter $\theta$
(a Rota--Baxter \textit{weight}) before the last term
of~\eqref{eq:Rota-Baxter}:
$$
K[S_1]\,K[S_2] = K\bigl[ K[S_1]S_2 + S_1K[S_2] - \theta S_1S_2 \bigr].
$$
The case $\theta = 0$ corresponds to a property of the integral
$I[f](x) := \int_0^x f(t)\,dt$, to wit,
\begin{equation}
I[f_1]\,I[f_2] = I\bigl[ I[f_1] f_2 + f_1 I[f_2] \bigr],
\label{eq:ach-so}
\end{equation}
which is just integration by parts. For $g$ fixed, the solution of the
equation $f = 1 - I[gf]$ with $I$ satisfying~\eqref{eq:ach-so} is
given by
$$
f = 1 - I[g] + I[gI[g]] - \cdots = e^{-I[g]},
$$
which follows from \eqref{eq:ach-so} for $f_1 = f_2 = f$, and
illustrates our approach in this paper.

Although~\eqref{eq:Rota-Baxter} is elementary, we wish to prove it
here. This is warranted because the Rota--Baxter property is
persistently ignored in field theory treatises; and this neglect is
not without consequences. For instance, in section 5.3.3 of the
standard text~\cite{Collins}, we find a tortured argument to try to
prove $\vft_-(\Ga_i \cup \Ga_j) = \vft_-(\Ga_i)\vft_-(\Ga_j)$, in
which the intermediate formulas (5.3.15) and (5.3.16) are plain wrong.
To see why \eqref{eq:Rota-Baxter} holds, notice that
\begin{align*}
K[S_1] S_2 + S_1 K[S_2] - S_1 S_2
&= K[S_1] (K[S_2] + K_+[S_2]) - (K[S_1] + K_+[S_1]) K_+[S_2]
\\
&= K[S_1] K[S_2] - K_+[S_1] K_+[S_2],
\end{align*}
and applying $K$ to this equality kills the term $K_+[S_1]\,K_+[S_2]$,
leaving $K[S_1]\,K[S_2]$ unchanged.

\section{The rationale for the matrix representation}

The not so mathematically-minded might wish to skip this section. In
the formalism Kreimer developed jointly with
Connes~\cite{ConnesKrI,ConnesKrII}, Feynman diagrams are organized in
a Hopf algebra $\H_\F$ of graphs; Feynman rules are understood as
linear and multiplicative maps $\vft$ of~$\H_\F$ into an algebra~$V$
(commutative, with unit) of quantum amplitudes; and the disentangling
of subdivergences is formulated as a factorization problem (Birkhoff
decomposition). The original version had a strong geometrical flavour,
but its supporting algebraic frame has emerged
since~\cite{EbrahimiFGuoKr}.

The space~$\H_\F$ is the algebra of polynomials with connected
Feynman graphs as indeterminates, multiplication being given by
simple juxtaposition of graphs. Connes and Kreimer introduced
on~$\H_\F$ a coproduct $\Delta : \H_\F \to \H_\F \ox \H_\F$,
serving to encode the superficially divergent subgraphs, by
setting $\Delta(\Ga) := \sum_{\Ga'} \Ga' \ox \Ga/\Ga'$, in the
notation of Section~\ref{sec:recipe}. For the coproducts of a Hopf
algebra~$\H$ one writes $\Delta a = \sum a_{(1)} \ox a_{(2)}$, for
$a \in \H$.

Let $(V,K)$ be a commutative Rota--Baxter algebra, and consider
$\Hom(\H,V)$, the space of linear maps from~$\H$ to~$V$; this is
an \textit{algebra} with the convolution operation, given by $f
\star g (a) = \sum f(a_{(1)}) \, g(a_{(2)})$, for $a \in \H$. In
our case the multiplicative (that is, product-respecting) elements
of $\Hom(\H_\F,V)$, with $V$ the algebra of Feynman amplitudes,
are of particular interest. Clearly they are determined by their
action on the subspace~$\F$ of connected graphs. We construct a
representation~$\Psi$ of~$\Hom(\H_\F,V)$ by infinite triangular
matrices with entries in $V$ by taking the composition
\begin{equation}
\Psi[f]: V\ox\F
\xrightarrow{\id_V\ox\Delta} V\ox \H_\F \ox \F
\xrightarrow{\id_V\ox f\ox\id_\F} V\ox V \ox \F
\xrightarrow{m_V\ox\id_\F} V\ox \F,
\end{equation}
where $m_V$ is just multiplication on~$V$. The plot works because the
external structure of the cographs $\Ga/\Ga'$ is the same as that
of~$\Ga$, so $\Delta$ actually sends $\F$ into $\H_\F\ox\F$. Thus for
any $f\in\Hom(\H_\F,V)$ a connected graph is sent by $\Psi[f]$ into a
linear combination of connected graphs with coefficients in~$V$,
corresponding to the same $n$-point function. In fact, $\Psi$ is an
antirepresentation, since $\Psi[f\star g] = \Psi[g]\Psi[f]$.

With the operator $\Krm$ given by $\Krm[f](a) := K[f(a)]$, the space
$\Hom(\H_\F,V)$ becomes a (noncommutative) Rota--Baxter algebra; then
$\Psi[\Krm[f]] = \K[\Psi[f]]$, with $\K$ the known matrix Rota--Baxter
map. Let finally $\vft \in \Hom(\H_\F,V)$ be the Feynman rule, which
is multiplicative, and denote
$$
\vf := \Psi[\vft].
$$
This will be a unipotent matrix. We have at last reproduced the
setting of this paper. The matrix
decomposition~\eqref{eq:mano-de-la-princesa} is a consequence of
Connes' and Kreimer's algebraic Birkhoff decomposition $\vft =
\vft_-^{-1} \star \vft_+$, where the two factors are multiplicative as
well~\cite{ConnesKrI}. Proofs and details are found in~\cite{Ausonia}.

\section{Conclusion}

Inspired by the Connes--Kreimer Hopf algebra formalism, we have
exhibited the combinatorics of renormalization as a collective
process, mechanized by means of simple matrix calculus. Our approach
neatly resolves the tension between the ``additive'' and the
``multiplicative'' sides of renormalization: the recursive
diagrammatic subtraction of subdivergences is the outcome of a
multiplicative process (this is not quite a trivial remark: a direct
check of the relation $\vft_+(\Ga) = K_+\Rbar(\Ga)$ if and only if
$\vft_-(\Ga) = -K_-\Rbar(\Ga)$ involves somewhat messy calculations).
As a consequence, the renormalization of the Lagrangian's parameters
by counterterms takes place by composition of series; the latter has
been known since 1855 to have a triangular matrix
representation~\cite{Bruna}. All this is more or less clear from the
analysis in~\cite{CaswellK,PivovarovT,ConnesKrII}; but probably
deserves further elucidation.

Also, we have rewritten the renormalization map in terms of the
projection $K_+$ on the finite part. In dimensional regularization
this prescription falls short of 't~Hooft's
desideratum~\cite{tHooftFinite}, as some of the terms in~$\vf_+$
contain coefficients of the pole parts; this objection, nevertheless,
loses force in regularization-free schemes like BPHZ and
Epstein--Glaser renormalization, that also possess a Rota--Baxter
property~\cite{Pomona}.

\subsection*{Acknowledgments}

KE-F is grateful to the Theory Department at the Physics Institute of
Bonn University for support. JMG-B thanks MEC, Spain, for support
through a `Ram\'on y Cajal' contract. LG is supported in part by NSF
grant DMS--0505643 and a grant from Rutgers University Research
Council. JCV acknowledges support from the Vicerrector\'{\i}a de
Investigaci\'on of the University of Costa~Rica.

\end{document}